\documentclass{article}

\usepackage{microtype}
\usepackage{graphicx}
\usepackage{subcaption}
\usepackage{booktabs} % for professional tables
\usepackage{hyperref}
\usepackage{multirow}
\usepackage{placeins}

\usepackage[preprint]{icml2026}

\usepackage{amsmath}
\usepackage{amssymb}
\usepackage{mathtools}
\usepackage{amsthm}

\usepackage[capitalize,noabbrev]{cleveref}

\usepackage[textsize=tiny]{todonotes}

\theoremstyle{plain}

\theoremstyle{definition}

\theoremstyle{remark}

\icmltitlerunning{
Adaptive Evidence Weighting for Audio-Spatiotemporal Fusion
}

\begin{document}

\twocolumn[
\icmltitle{
% Adaptive Log-Linear Evidence Fusion for Decision-Theoretically Safe Posteriors
% Adaptive Evidence Fusion Models for Bioacoustic Classification
% Adaptive Evidence Fusion for Joint Audio-Timeseries Models
%Adaptive Evidence Fusion for Joint Audio-Spatio Temporal Modeling
%Adaptive Evidence Fusion for Audio-Spatiotemporal Classification
Adaptive Evidence Weighting for Audio-Spatiotemporal Fusion
}

% Anonymous submission: DO NOT include author info.
% The ICML style file will suppress author blocks in review mode anyway, but
% keep this template clean.

\icmlkeywords{bioacoustics, evidence fusion, priors, uncertainty, ecological context}

\icmlsetsymbol{equal}{*}
\begin{icmlauthorlist}
\icmlauthor{Oscar Ovanger}{ut}
\icmlauthor{Levi Harris}{unc}
\icmlauthor{Timothy H. Keitt}{ut}
\end{icmlauthorlist}
\icmlaffiliation{ut}{Jackson School of Geosciences, University of Texas at Austin, Austin, TX, USA}
\icmlcorrespondingauthor{Oscar Ovanger}{oscar.ovanger@utexas.edu}
\printAffiliationsAndNotice{}

\vskip 0.3in
]

% "A general framework for fusing classifers."

% Required even if empty:

%%%%%%%%%%%%%%%%%%%%%%%%%%%%%%%%%%%%%%%%%%%%%%%%%%%%%%%%%%%%%%%%%%%%%%%%%%%%%%%
% ABSTRACT (already in good shape; keep as single paragraph 4--6 sentences)
%%%%%%%%%%%%%%%%%%%%%%%%%%%%%%%%%%%%%%%%%%%%%%%%%%%%%%%%%%%%%%%%%%%%%%%%%%%%%%%

\begin{abstract}
Many machine learning systems have access to multiple sources of evidence for the same prediction target, yet these sources often differ in reliability and informativeness across inputs. In bioacoustic classification, species identity may be inferred both from the acoustic signal and from spatiotemporal context such as location and season; while Bayesian inference motivates multiplicative evidence combination, in practice we typically only have access to discriminative predictors rather than calibrated generative models.
We introduce \textbf{F}usion under \textbf{IN}dependent \textbf{C}onditional \textbf{H}ypotheses (\textbf{FINCH}), an adaptive log-linear evidence fusion framework that integrates a pre-trained audio classifier with a structured spatiotemporal predictor.
FINCH learns a per-sample gating function that estimates the reliability of contextual information from uncertainty and informativeness statistics.
The resulting fusion family \emph{contains} the audio-only classifier as a special case and explicitly bounds the influence of contextual evidence, yielding a risk-contained hypothesis class with an interpretable audio-only fallback.
Across benchmarks, FINCH consistently outperforms fixed-weight fusion and audio-only baselines, improving robustness and error trade-offs even when contextual information is weak in isolation.
We achieve state-of-the-art performance on CBI and competitive or improved performance on several subsets of BirdSet using a lightweight, interpretable, evidence-based approach.
Code is available:
\texttt{\href{https://anonymous.4open.science/r/birdnoise-85CD/README.md}{anonymous-repository}}
\end{abstract}

%%%%%%%%%%%%%%%%%%%%%%%%%%%%%%%%%%%%%%%%%%%%%%%%%%%%%%%%%%%%%%%%%%%%%%%%%%%%%%%
% 1 INTRODUCTION
%%%%%%%%%%%%%%%%%%%%%%%%%%%%%%%%%%%%%%%%%%%%%%%%%%%%%%%%%%%%%%%%%%%%%%%%%%%%%%%
\begin{figure}[htb]
    \centering
    \includegraphics[width=0.5\linewidth]{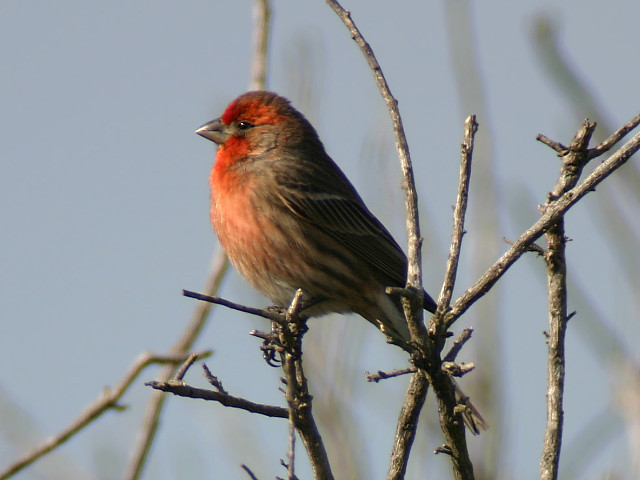}
    \caption{\textit{Haemorhous mexicanus} (North American House Finch), the namesake of our model. }
    \label{fig:placeholder}
\end{figure}
\section{Introduction}
\label{sec:intro}

Ensemble learning studies how multiple predictive models for the same target variable can be combined to improve performance, robustness, or calibration relative to any single model \cite{kuncheva2004combining,kittler1998combining}. Classical ensemble methods such as bagging, boosting, and stacking typically assume that all predictors are trained on the same underlying data distribution and that ensemble weights or combination rules can be learned jointly with model parameters.

In many modern applications, however, predictive models are pre-trained, heterogeneous, and fixed at inference time. These models may rely on different sources of evidence, be trained on distinct datasets, and operate under incompatible modeling assumptions. As a result, joint retraining or fine-tuning is often infeasible. Instead, the problem becomes one of combining the outputs of fixed predictors in a principled manner.

We consider the common setting in which multiple predictive models provide complementary evidence about a shared latent target.
Let $y$ denote a class label, and let $x$ and $s$ denote two sources of evidence.
Conditional independence,
\begin{equation}
x \perp s \mid y,
\end{equation}
provides a useful idealization that motivates multiplicative evidence fusion.
If the full generative distributions were available, Bayesian inference would yield
\begin{equation}
p(y \mid x, s) \propto p(x \mid y)\, p(s \mid y)\, p(y).
\end{equation}
FINCH adopts the corresponding log-linear fusion form while remaining applicable in regimes where this independence assumption holds only approximately, and deviations can be mitigated through adaptive, bounded weighting.

In practice, however, generative models are often unavailable. Instead, one typically has access only to discriminative predictors $p_\theta(y \mid x)$ and $p_\psi(y \mid s)$, trained independently. In this case, the posterior can be expressed as
\begin{equation}
p(y \mid x, s) \propto \frac{p_\theta(y \mid x)\, p(x)\, p_\psi(y \mid s)\, p(s)}{p(y)},
\end{equation}
where the marginal distributions $p(x)$, $p(s)$, and the implied prior $p(y)$ are generally unknown. As a result, the true posterior is not directly computable from the discriminative models alone.

A common approximation in this setting is to combine discriminative predictors using log-linear models or product-of-experts formulations \cite{hinton2002products,genest1986combining}. These models operate directly on posterior outputs and define a fused distribution of the form
\begin{equation}
\log p(y \mid x, s) = \log p_\theta(y \mid x) + \log p_\psi(y \mid s) - \log Z(x,s),
\end{equation}
where $Z(x,s)$ is a normalization constant ensuring $\sum_y p(y\mid x,s)=1$. Throughout the paper, we write fusion rules in unnormalized log-space; the normalization constant is implicitly handled by the final softmax and is therefore omitted when it does not affect comparisons across classes. Log-linear fusion is decision-theoretically justified as a logarithmic opinion pool under mild axioms \cite{heskes1998logopinion} and provides a practical surrogate when only discriminative models are available.

Existing approaches typically employ fixed or globally learned weights when combining predictors. This implicitly assumes that the relative reliability of each evidence source is constant across the input space. In many applications, this assumption does not hold. The informativeness of a given evidence source may vary substantially across samples, leading to degraded performance or pathological dominance when fixed weights are used.

In this work, we introduce an adaptive log-linear fusion framework that preserves the structure and interpretability of classical product-of-experts models while allowing per-sample modulation of evidence strength.
We assume that the constituent predictors are fixed and discriminative, and we do not retrain or recalibrate their parameters.
Instead, we learn a gating function that estimates the reliability of contextual evidence on a per-sample basis.
The resulting fused posterior is defined (up to normalization) as
\begin{equation}
\label{eq:adaptive_fusion}
\log \tilde{p}_\omega(y \mid x, s)
=
\log p_\theta(y \mid x)
+
\omega(x,s)\, \log p_\psi(y \mid s),
\end{equation}
where $\omega(x,s)\ge 0$ is a learned weighting function.
The normalized posterior $p_\omega(y\mid x,s)$ is obtained by applying a softmax over $y$. This formulation recovers the first model classifier when $\omega(x,s)=0$, bounds the influence of contextual evidence, and enables adaptive fusion without retraining the base models.

We evaluate this framework in the context of bioacoustic species classification by combining a state-of-the-art acoustic classifier with a structured spatiotemporal prior derived from large-scale observational data. Experiments show that adaptive weighting consistently outperforms fixed-weight fusion and audio-only baselines, particularly in regimes where contextual information is heterogeneous or weak in isolation.

%%%%%%%%%%%%%%%%%%%%%%%%%%%%%%%%%%%%%%%%%%%%%%%%%%%%%%%%%%%%%%%%%%%%%%%%%%%%%%%
% 2 RELATED WORK
%%%%%%%%%%%%%%%%%%%%%%%%%%%%%%%%%%%%%%%%%%%%%%%%%%%%%%%%%%%%%%%%%%%%%%%%%%%%%%%

\section{Related Work}
\label{sec:related}

We organize related work into two themes: (i) \emph{theory and mechanisms} for combining probabilistic predictors (log-linear pooling, products of experts, and gating), and (ii) \emph{bioacoustic applications}, distinguishing audio-only foundation encoders from systems that incorporate spatiotemporal context.

\subsection{Log-Linear Fusion, Opinion Pools, and Products of Experts}
The combination of multiple classifiers has been studied extensively in machine learning and statistics \cite{kuncheva2004combining,kittler1998combining}. Classical ensemble methods (e.g., averaging, voting, Bayesian combination) depend on classifier diversity and error correlation, and often learn weights jointly with model parameters or via validation tuning.

Maximum entropy models and logarithmic opinion pools formalize classifier fusion as a log-linear aggregation of predictive distributions \cite{berger1996maximum,genest1986combining}. Product-of-experts models combine distributions by multiplication, yielding sharp posteriors when experts agree and diffuse posteriors when they disagree \cite{hinton2002products}. Logarithmic opinion pools provide a decision-theoretic justification for weighted log-linear combinations under mild axioms \cite{heskes1998logopinion}, and log-linear fusion can be viewed as a practical surrogate to Bayesian inference when only discriminative predictors are available \cite{bishop2006prml}.

Most prior approaches employ fixed or globally learned weights, implicitly assuming that expert reliability is stationary across the input space. Our work focuses instead on \emph{per-sample} weighting of contextual evidence while preserving the interpretability of log-linear fusion.

\subsection{Gating Networks, Mixtures of Experts, and Reliability Estimation}
Mixture-of-experts models introduce gating networks that assign input-dependent weights to multiple predictors \cite{jacobs1991adaptive,jordan1994hierarchical}. These models are typically trained end-to-end, with both experts and gates optimized jointly.

Our setting differs in two key respects. First, we treat the constituent predictors as fixed and trained independently on different evidence sources. Second, we use \emph{multiplicative} fusion in probability space (log-additive in log-space), motivated by conditional independence of evidence sources given the label. The gating function modulates the \emph{strength} of contextual evidence rather than selecting among experts.

A related line of work studies uncertainty estimation and selective prediction. Bayesian approximations and calibrated uncertainty signals \cite{gal2016dropout,kendall2017uncertainties} as well as confidence-based methods \cite{hendrycks2017baseline} are commonly used to identify unreliable predictions or out-of-distribution inputs. Selective classification allows abstention when confidence is low \cite{geifman2017selective}. In contrast, our method uses reliability signals to continuously modulate the contribution of contextual evidence within a probabilistic fusion model, rather than abstaining or discarding an expert.

\subsection{Audio Foundation Models for Bioacoustics (Audio-Only Encoders)}
Modern bioacoustic classifiers increasingly rely on self-supervised or large-scale pretrained audio encoders, followed by lightweight adaptation (e.g., linear probes or prototypical probes) on downstream datasets \cite{Lauha2022,Wang2022,hagiwara2022avesanimalvocalizationencoder,Choudhary_2022,Nolasco2023,rauch2025birdsetlargescaledatasetaudio}.
Recent foundation-model style approaches provide strong transferable representations for animal sounds, including BEATs \cite{chen2022} and NatureLM-Audio \cite{robinson2025}, as well as large-scale benchmark/analysis efforts such as BEANS and studies of what drives bioacoustic representation quality \cite{hagiwara2022beansbenchmarkanimalsounds,miron2025mattersbioacousticencoding}.
Perch-style models emphasize large-scale pretraining for bird and wildlife acoustics and provide strong embeddings for transfer learning \cite{ghani2023perch,perchiipointzero}.

These audio-only encoders are powerful because they can be pretrained on audio corpora without requiring spatiotemporal metadata, and then reused across tasks and regions. Our method explicitly preserves this modularity by freezing the audio encoder/classifier and incorporating context only at the fusion layer.

\subsection{Spatiotemporal Context in Bioacoustics: Priors vs.\ Joint Models}
Contextual information such as location and season is often incorporated via ecological priors derived from species distribution models \cite{fink2013adaptive,Madhusudhana2021}, including large-scale citizen-science abundance models \cite{fink2010spatiotemporal,Fink2014,Jonhston2015}. Many practical bioacoustic systems combine acoustic predictions with such priors either heuristically (e.g., post-hoc filtering) or by concatenating metadata with learned audio features.

BirdNET is a widely used system for avian monitoring and is commonly deployed with metadata-informed constraints (e.g., location/date based filtering or contextual feasibility) alongside acoustic classification \cite{kahl2021birdnet}. More generally, joint approaches that ingest audio and metadata into a single predictor $p(y\mid x,s)$ are more expressive but must learn audio-context interactions directly and require paired supervision. As formalized in Appendix~A2, such joint modeling becomes necessary when conditional dependence between evidence sources is strong, whereas under (approximate) conditional independence log-linear fusion is sufficient and avoids the complexity of retraining large multimodal architectures.

Our contribution targets the common regime where context is informative but heterogeneous: we combine a strong pretrained audio classifier with a structured spatiotemporal prior through a bounded, per-sample gating function, improving robustness while preserving audio-only fallback and modular pretraining benefits.

%%%%%%%%%%%%%%%%%%%%%%%%%%%%%%%%%%%%%%%%%%%%%%%%%%%%%%%%%%%%%%%%%%%%%%%%%%%%%%%
% 3 METHOD
%%%%%%%%%%%%%%%%%%%%%%%%%%%%%%%%%%%%%%%%%%%%%%%%%%%%%%%%%%%%%%%%%%%%%%%%%%%%%%%
\section{Method}
\label{sec:method}

\subsection{Notation and Evidence Sources}
\label{sec:notation}

We consider a multi-class classification problem with the following notation:
\begin{itemize}
    \item $x \in \mathcal{X}$: audio observation, represented as a 3\,s log-mel spectrogram,
    \item $s \in \mathcal{S}$: spatiotemporal context (latitude, longitude, and time),
    \item $y \in \mathcal{Y}$: species label,
    \item $p_\theta(y\mid x)$: audio-based classifier,
    \item $p_\psi(y\mid s)$: spatiotemporal (ecological) classifier,
    \item $p_\omega(y\mid x,s)$: fused classifier.
\end{itemize}

While our application is bioacoustic classification, the proposed fusion framework applies generally to any setting with multiple conditionally independent evidence sources given the target label.

\subsection{Log-Linear Fusion Family}
\label{sec:fusion}

We adopt a per-sample log-linear fusion model of the form

\begin{align}
\label{eq:fusion}
\log \tilde{p}_{\omega}(y\mid x,s)
&= \log p_{\theta}(y\mid x)
+ \omega(x,s)\,\log p_{\psi}(y\mid s),
\end{align}
where $\tilde{p}_{\omega}$ denotes an unnormalized log-score. The normalized posterior
$p_{\omega}(y\mid x,s)$ is obtained by applying a softmax over $y$. The scalar $\omega(x,s)\ge 0$ controls the influence of the spatiotemporal model on a per-sample basis. We parameterize $\omega(x,s)$ using a bounded transformation with learnable scale:
\[
\omega(x,s) = \omega_{\max}\,\sigma\!\big(g_\phi(u(x,s))\big) + \epsilon,
\]
where $\sigma(\cdot)$ denotes the sigmoid function, $\epsilon>0$ ensures numerical stability, and $\omega_{\max}$ is a trainable parameter constrained to lie in $[\epsilon,10]$.
This design allows the model to adaptively learn the appropriate range of contextual influence from data, while preventing degenerate or unbounded fusion weights.
When $\omega(x,s)=0$, the fused classifier exactly recovers the audio-only model $p_\theta(y\mid x)$.

The fusion is intentionally asymmetric: the audio classifier contributes additively in log-space for all samples, whereas the contextual model modulates the posterior only when deemed informative. This reflects the complementary inductive biases of the two sources. Audio observations often provide high discriminative power but are susceptible to confusions between acoustically similar species, while spatiotemporal models provide coarse but robust constraints by encoding ecological feasibility.

\subsection{Adaptive Gating: Reliability and Informativeness}
\label{sec:gating}

To adaptively weight contextual evidence, we learn a gating function $\omega(x,s)$ parameterized by a two-layer MLP. The gating network operates on a feature vector $u(x,s)$ that summarizes uncertainty, confidence, and contextual structure:

% \subsubsection{Audio features}
% \[
%     \scalebox{0.8}{$\displaystyle
%     f_{\text{audio}}(x) =
%     \left[
%         \max_y p_\theta(y\mid x),\;
%         H\!\left(p_\theta(y\mid x)\right),\;
%         p_\theta(y_1\mid x)-p_\theta(y_2\mid x)
%     \right],
%     $}
% \]
\begin{itemize}
    \item \textbf{Audio features}:
    \[
    \scalebox{0.8}{$\displaystyle
    f_{\text{audio}}(x) =
    \left[
        \max_y p_\theta(y\mid x),\;
        H\!\left(p_\theta(y\mid x)\right),\;
        p_\theta(y_1\mid x)-p_\theta(y_2\mid x)
    \right],
    $}
    \]
\end{itemize}

% \begin{itemize}
% \item \textbf{Audio features}:
% \begin{multline*}
% f_{\text{audio}}(x) = \bigl[
%     \max_y p_\theta(y\mid x),\;
%     H\!\left(p_\theta(y\mid x)\right),\\
%     p_\theta(y_1\mid x)-p_\theta(y_2\mid x)
% \bigr]
% \end{multline*}

where $y_1,y_2$ denote the top two predicted classes.

    % \item \textbf{Spatiotemporal features}:
    % \[
    % f_{\text{prior}}(s) =
    % \left[
    %     \max_y p_\psi(y\mid s),\;
    %     H\!\left(p_\psi(y\mid s)\right),\;
    %     p_\psi(y_1\mid s)-p_\psi(y_2\mid s)
    % \right].
    % \]

% \subsubsection{Spatiotemporal features}
% \[
%     \scalebox{0.8}{$\displaystyle
%     f_{\text{prior}}(s) =
%     \left[
%         \max_y p_\psi(y\mid s),\;
%         H\!\left(p_\psi(y\mid s)\right),\;
%         p_\psi(y_1\mid s)-p_\psi(y_2\mid s)
%     \right].
%     $}
% \]

\begin{itemize}
    \item \textbf{Spatiotemporal features}:
    \[
    \scalebox{0.8}{$\displaystyle
    f_{\text{prior}}(s) =
    \left[
        \max_y p_\psi(y\mid s),\;
        H\!\left(p_\psi(y\mid s)\right),\;
        p_\psi(y_1\mid s)-p_\psi(y_2\mid s)
    \right].
    $}
    \]
\end{itemize}
\begin{itemize}
    \item \textbf{Metadata features}:
    \[
    \scalebox{0.85}{$\displaystyle
    f_{\text{meta}} =
    \left[
        \sin\!\left(\tfrac{2\pi d}{365}\right),
        \cos\!\left(\tfrac{2\pi d}{365}\right),
        \sin\!\left(\tfrac{2\pi h}{24}\right),
        \cos\!\left(\tfrac{2\pi h}{24}\right),
        \tfrac{\text{lat}}{90},
        \tfrac{\text{lon}}{180}
    \right].
    $}
    \]
\end{itemize}
    % \[
    % \scalebox{0.8}{$\displaystyle
    % f_{\text{meta}} =
    % \left[
    %     \sin\!\left(\tfrac{2\pi d}{365}\right),
    %     \cos\!\left(\tfrac{2\pi d}{365}\right),
    %     \sin\!\left(\tfrac{2\pi h}{24}\right),
    %     \cos\!\left(\tfrac{2\pi h}{24}\right),
    %     \tfrac{\text{lat}}{90},
    %     \tfrac{\text{lon}}{180}
    % \right].
    % \]
    % \[

% \subsubsection{Metadata features}
% \begin{itemize}
%     \[
%     f_{\text{meta}} =
%     \left[
%         \sin\!\left(\tfrac{2\pi d}{365}\right),
%         \cos\!\left(\tfrac{2\pi d}{365}\right),
%         \sin\!\left(\tfrac{2\pi h}{24}\right),
%         \cos\!\left(\tfrac{2\pi h}{24}\right),
%         \tfrac{\text{lat}}{90},
%         \tfrac{\text{lon}}{180}
%     \right].
%     \]
%     \[
%     \scalebox{0.85}{$\displaystyle
%     f_{\text{meta}} =
%     \left[
%         \sin\!\left(\tfrac{2\pi d}{365}\right),
%         \cos\!\left(\tfrac{2\pi d}{365}\right),
%         \sin\!\left(\tfrac{2\pi h}{24}\right),
%         \cos\!\left(\tfrac{2\pi h}{24}\right),
%         \tfrac{\text{lat}}{90},
%         \tfrac{\text{lon}}{180}
%     \right].
%     $}
%     \]
% \end{itemize}

The concatenated feature vector $u(x,s)$ is passed through a two-layer MLP $g_\phi(u)$ with ReLU activations and dropout. Its output is squashed through a sigmoid and scaled by $\omega_{\max}$, yielding $\omega(x,s)\in[0,\omega_{\max}]$.

In addition to the gating parameters, we jointly learn a temperature parameter $T$ applied to the audio logits, and introduce a small constant $\epsilon$ to ensure numerical stability. The resulting fused model is

\begin{align}
\label{eq:fused_temp}
\log \tilde{p}(y\mid x,s)
&= \frac{\log p_\theta(y\mid x)}{T}
+ \omega(x,s)\log\!\big(p_\psi(y\mid s)+\epsilon\big),
\end{align}
and we obtain $p(y\mid x,s)$ by applying a softmax over $y$.

\paragraph{Avoiding Gate Collapse.}
To prevent the adaptive gate from collapsing to a trivial solution (e.g., $\omega(x,s)\equiv 0$), we introduce a variance regularization term during training:
\[
\mathcal{L}_{\mathrm{var}} = -\lambda_{\mathrm{var}}\,\mathrm{Var}_{(x,s)\sim\mathcal{B}}[\omega(x,s)],
\]
where the variance is computed over each minibatch $\mathcal{B}$. This term encourages $\omega(x,s)$ to take non-constant values across inputs, promoting genuine adaptivity rather than global suppression of contextual evidence.
We set $\lambda_{\mathrm{var}}$ sufficiently small and rely on validation-based model selection so that the regularizer promotes non-trivial adaptivity without overriding the empirical preference for $\omega(x,s)\approx 0$ when context is helpful.

\subsection{Decision-Theoretic Safety and Robustness}
\label{sec:safety}

The proposed fusion framework is designed to be robust to misspecified or weak contextual models.
A key structural property is recoverability: for any input $(x,s)$, setting $\omega(x,s)=0$ exactly recovers the audio-only classifier.
Thus, the fusion family contains an explicit audio-only fallback and does not require reliance on contextual information.

The influence of the spatiotemporal model is further controlled by bounding the fusion weight $\omega(x,s)\le \omega_{\max}$, which prevents pathological domination of the posterior by contextual evidence.
Even highly confident but incorrect contextual predictions therefore cannot overwhelm strong acoustic evidence.

Adaptive gating enables selective incorporation of context.
When the spatiotemporal model is uninformative (e.g., near-uniform or high entropy), the learned gate is driven toward zero.
When contextual evidence is informative and consistent with the audio prediction, increasing $\omega(x,s)$ sharpens the posterior by suppressing ecologically implausible alternatives.

From a decision-theoretic perspective, the fused classifier corresponds to minimizing expected log-loss under a bounded, sample-dependent combination of log-likelihood terms.
Recoverability guarantees that the audio-only hypothesis remains available within the model class, while bounded gating mitigates harmful over-reliance on contextual evidence in practice.
Formal conditions are provided in the appendix.

Finally, freezing the audio classifier throughout fusion training preserves the underlying acoustic decision function, ensuring that any changes in prediction arise solely from the fusion mechanism rather than modification of the audio model itself.

%%%%%%%%%%%%%%%%%%%%%%%%%%%%%%%%%%%%%%%%%%%%%%%%%%%%%%%%%%%%%%%%%%%%%%%%%%%%%%%
% FIGURE 1 (LOCKED): Schematic of method
%%%%%%%%%%%%%%%%%%%%%%%%%%%%%%%%%%%%%%%%%%%%%%%%%%%%%%%%%%%%%%%%%%%%%%%%%%%%%%%
\begin{figure*}[htb]
  \centering
  \includegraphics[width=\linewidth]{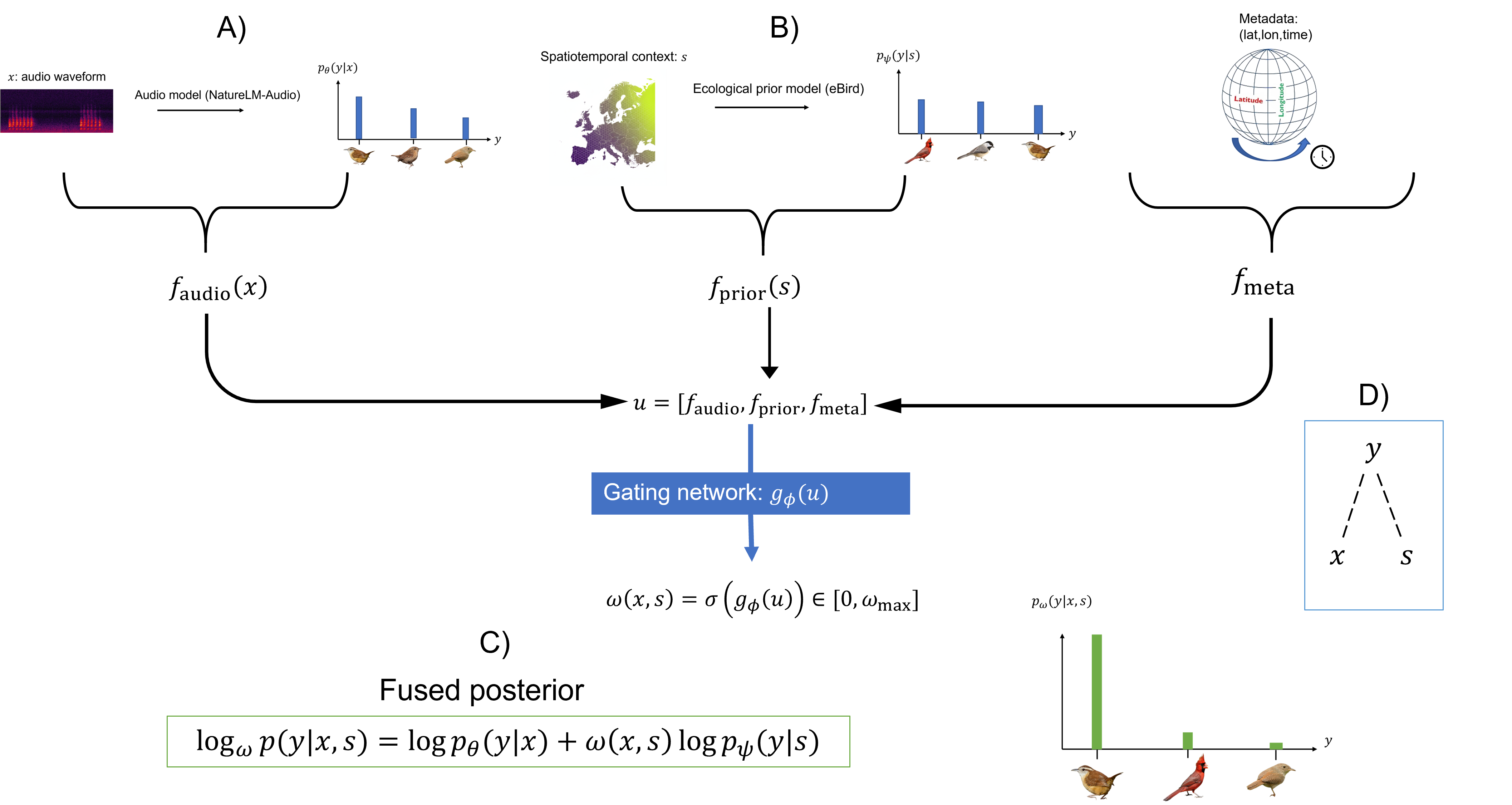}
  \caption{
  \textbf{FINCH: adaptive log-linear evidence fusion.}
  \textbf{(A)} An audio encoder and linear head produce an acoustic posterior $p_\theta(y\mid x)$ from the input spectrogram.
  \textbf{(B)} A spatiotemporal model maps context $s$ (location and time) to a prior $p_\psi(y\mid s)$.
  \textbf{(C)} A gating network computes a nonnegative per-sample weight $\omega(x,s)$ from uncertainty and informativeness features, yielding unnormalized log-scores
$\log \tilde{p}_\omega(y\mid x,s)=\log p_\theta(y\mid x)+\omega(x,s)\log p_\psi(y\mid s)$,
which are normalized by a softmax over $y$ to obtain $p_\omega(y\mid x,s)$.
  When $\omega(x,s)=0$, the audio-only model is recovered.
  \textbf{(D)} Directed acyclic graph illustrating the assumption that $x$ and $s$ are conditionally independent given the label $y$, motivating multiplicative fusion.
  }
  \label{fig:schematic}
\end{figure*}

%%%%%%%%%%%%%%%%%%%%%%%%%%%%%%%%%%%%%%%%%%%%%%%%%%%%%%%%%%%%%%%%%%%%%%%%%%%%%%%
% 4 EXPERIMENTAL SETUP
%%%%%%%%%%%%%%%%%%%%%%%%%%%%%%%%%%%%%%%%%%%%%%%%%%%%%%%%%%%%%%%%%%%%%%%%%%%%%%%
\section{Experimental Setup}
\label{sec:expsetup}

\subsection{Datasets and Spatiotemporal Prior}
\label{subsec:data_prior}

We evaluate our method on two large-scale bioacoustic benchmarks: the Cornell Birdcall Identification (CBI) dataset and BirdSet. 
CBI is derived from the eBird citizen-science corpus and was released as part of a Kaggle competition \cite{cbi_data}. The dataset contains short audio recordings annotated with species labels and associated metadata, including latitude, longitude, and date. In total, CBI comprises 264 bird species.
BirdSet is a large-scale benchmark for avian audio classification with standardized splits and evaluation protocols \cite{rauch2025birdsetlargescaledatasetaudio}.

We incorporate weekly species-level spatiotemporal abundance priors from the eBird Status and Trends project (AdaSTEM) \cite{fink2010spatiotemporal,Fink2014,Jonhston2015}, queried by location and date. Priors are precomputed for training efficiency and used directly at inference. Full details are provided in Appendix~A5.

\paragraph{Context model per benchmark.}
We use different spatiotemporal models depending on the dataset. On CBI, $p_\psi(y\mid s)$ is given by the external eBird Status \& Trends (AdaSTEM) prior queried by location and date (Appendix~A5). On BirdSet, AdaSTEM priors are not available; instead, we train a lightweight metadata-only MLP on the BirdSet training split to obtain a learned contextual predictor $p_\psi(y\mid s)$ from the same time/location encodings used by the gating network (Appendix~A1).

\subsection{Audio Encoder and Classification Head}
\label{subsec:audio_model}

For audio representation learning, we use the BEATs encoder extracted from the NatureLM-Audio model \cite{robinson2025}. Although NatureLM-Audio is trained end-to-end for audio-language tasks, prior work has shown that the resulting BEATs encoder achieves state-of-the-art performance on bioacoustic benchmarks when used as a frozen feature extractor \cite{miron2025mattersbioacousticencoding}.

All parameters of the audio encoder are frozen throughout training. The encoder produces embeddings of dimension $4096$, which are mean-pooled over time and passed to a linear classification head
\begin{equation}
y = A h(x) + b,
\end{equation}
where $A \in \mathbb{R}^{C \times 4096}$ and $b \in \mathbb{R}^{C}$, with $C$ denoting the number of species. For CBI, $C = 264$.

\subsection{Fusion Models and Training Procedure}
\label{subsec:fusion_models}

To disentangle representation learning from evidence fusion, we adopt a three-stage training procedure in which the audio encoder remains frozen throughout.

\paragraph{Stage 1: Audio-Only Training.}
We first train a linear classifier on frozen audio embeddings to obtain an audio-only predictor $p_\theta(y \mid x)$, establishing a strong acoustic baseline.

\paragraph{Stage 2: Fixed-Weight Fusion.}
We then introduce a fixed scalar fusion weight $\omega \ge 0$, shared across all samples, to combine audio and spatiotemporal evidence. This stage learns a globally calibrated fusion baseline.

\paragraph{Stage 3: Adaptive Gating Fusion.}
Finally, we replace the global scalar weight with an adaptive gating network (Section~\ref{sec:gating}) that predicts a nonnegative, sample-dependent fusion weight $\omega(x,s)$. The gating network is initialized to match the fixed-weight solution and trained to enable per-sample adaptivity.

In Table~\ref{tab:training_setup} is a summary of the model components and trainable parameters for each training stage.

%%%%%%%%%%%%%%%%%%%%%%%%%%%%%%%%%%%%%%%%%%%%%%%%%%%%%%%%%%%%%%%%%%%%%%%%%%%%%%%
% 5 Results
%%%%%%%%%%%%%%%%%%%%%%%%%%%%%%%%%%%%%%%%%%%%%%%%%%%%%%%%%%%%%%%%%%%%%%%%%%%%%%%

\section{Results}
\label{sec:results}

\subsection{Main Benchmark Results (Aggregate)}
\label{sec:mainresults}

\begin{table*}[t]
\centering
\caption{
\textbf{Main results on CBI and BirdSet subsets.}
For BirdSet, we report retrieval AUROC (R), detection cmAP (m), and Top-1 accuracy (A) on four subsets (PER, NES, UHH, SSW), formatted as \textit{R / m / A}.
For CBI, we report Top-1 accuracy and the corresponding training regime (“Method”).
FINCH uses an external AdaSTEM prior on CBI and a learned metadata-only MLP prior on BirdSet.
}
\label{tab:BirdSET}
\setlength{\tabcolsep}{3.5pt}
\renewcommand{\arraystretch}{1.05}
\scriptsize
\begin{tabular*}{\textwidth}{@{\extracolsep{\fill}} lcc cccc}
\toprule
\multirow{2}{*}{Model}
& \multicolumn{2}{c}{\textbf{CBI}}
& \multicolumn{4}{c}{\textbf{BirdSet (subset metrics: R / m / A)}} \\
\cmidrule(lr){2-3}\cmidrule(lr){4-7}
& Method & Acc
& PER (R/m/A) & NES (R/m/A) & UHH (R/m/A) & SSW (R/m/A) \\
\midrule

Audio ProtoPNet-5
& -- & --
& 0.790 / 0.300 / 0.590
& 0.930 / 0.380 / 0.520
& 0.870 / 0.310 / 0.490
& 0.970 / 0.420 / 0.660 \\

BirdMAE-L
& -- & --
& \underline{0.820} / \textbf{0.350} / \underline{0.600}
& 0.910 / \textbf{0.410} / 0.520
& 0.820 / 0.300 / 0.420
& 0.930 / 0.410 / 0.700 \\

BirdMAE-L (PP)
& -- & --
& \underline{0.820} / 0.310 / 0.590
& 0.930 / 0.380 / 0.470
& 0.830 / 0.300 / 0.360
& 0.940 / 0.380 / 0.620 \\

Perch 2.0 -- Peak-select
& LP & 0.785
& 0.802 / 0.255 / 0.528
& \underline{0.948} / 0.382 / 0.504
& 0.901 / 0.375 / 0.462
& \textbf{0.974} / \underline{0.455} / \underline{0.782} \\

Perch 2.0 -- Random
& LP & 0.792
& 0.786 / 0.232 / 0.535
& \textbf{0.953} / \underline{0.403} / 0.562
& \underline{0.912} / \underline{0.380} / 0.595
& \underline{0.973} / \textbf{0.469} / \textbf{0.789} \\

\midrule
Audio-only ($p_\theta(y\mid x)$)
& LP & \underline{0.806}
& -- & -- & -- & -- \\

Prior-only ($p_\psi(y\mid s)$)
& LP & 0.030
& -- & -- & -- & -- \\

FINCH (ours)
& LP & \textbf{0.826}
& \textbf{0.824} / 0.232 / 0.429
& 0.936 / 0.245 / \textbf{0.679}
& \textbf{0.927} / \textbf{0.536} / \textbf{0.747}
& 0.642 / 0.025 / 0.688 \\

\bottomrule
\end{tabular*}
\end{table*}

Table~\ref{tab:benchmark} summarizes performance on CBI and BirdSet subsets, comparing FINCH against strong audio-only baselines and existing bioacoustic systems.
For CBI, FINCH uses an external spatiotemporal prior derived from eBird Status \& Trends (AdaSTEM), whereas for BirdSet it uses a learned metadata-only MLP prior trained on the BirdSet training split.

\paragraph{CBI.}
On CBI, FINCH achieves the highest test accuracy among all evaluated methods under the linear-probe protocol, improving accuracy from $0.806$ for the audio-only baseline to $0.826$ with adaptive fusion.
Fixed-weight fusion yields only marginal improvement ($0.808$), indicating that globally calibrated fusion is insufficient when contextual reliability varies across inputs.
The spatiotemporal prior performs poorly in isolation (accuracy $=0.030$), confirming that the observed gains arise from selective integration rather than reliance on context alone.
Despite introducing only a lightweight gating network on top of frozen predictors, FINCH outperforms both audio-only models and joint audio–context systems evaluated under the same protocol.

\paragraph{BirdSet.}
On BirdSet subsets, FINCH matches or improves upon strong audio-only baselines across multiple evaluation metrics, including retrieval (AUROC), detection (cmAP), and classification (Top-1 accuracy).
These results are notable given that the contextual predictor is a simple metadata-only MLP trained from limited supervision.
While improvements vary across subsets and metrics, FINCH consistently demonstrates that adaptive evidence weighting can leverage weak and heterogeneous contextual signals without degrading performance relative to audio-only models.

\subsection{Qualitative Analysis: Adaptive Fusion in Practice}
\label{sec:qualitative}

\begin{figure*}[htb]
  \centering
  \includegraphics[width=\linewidth]{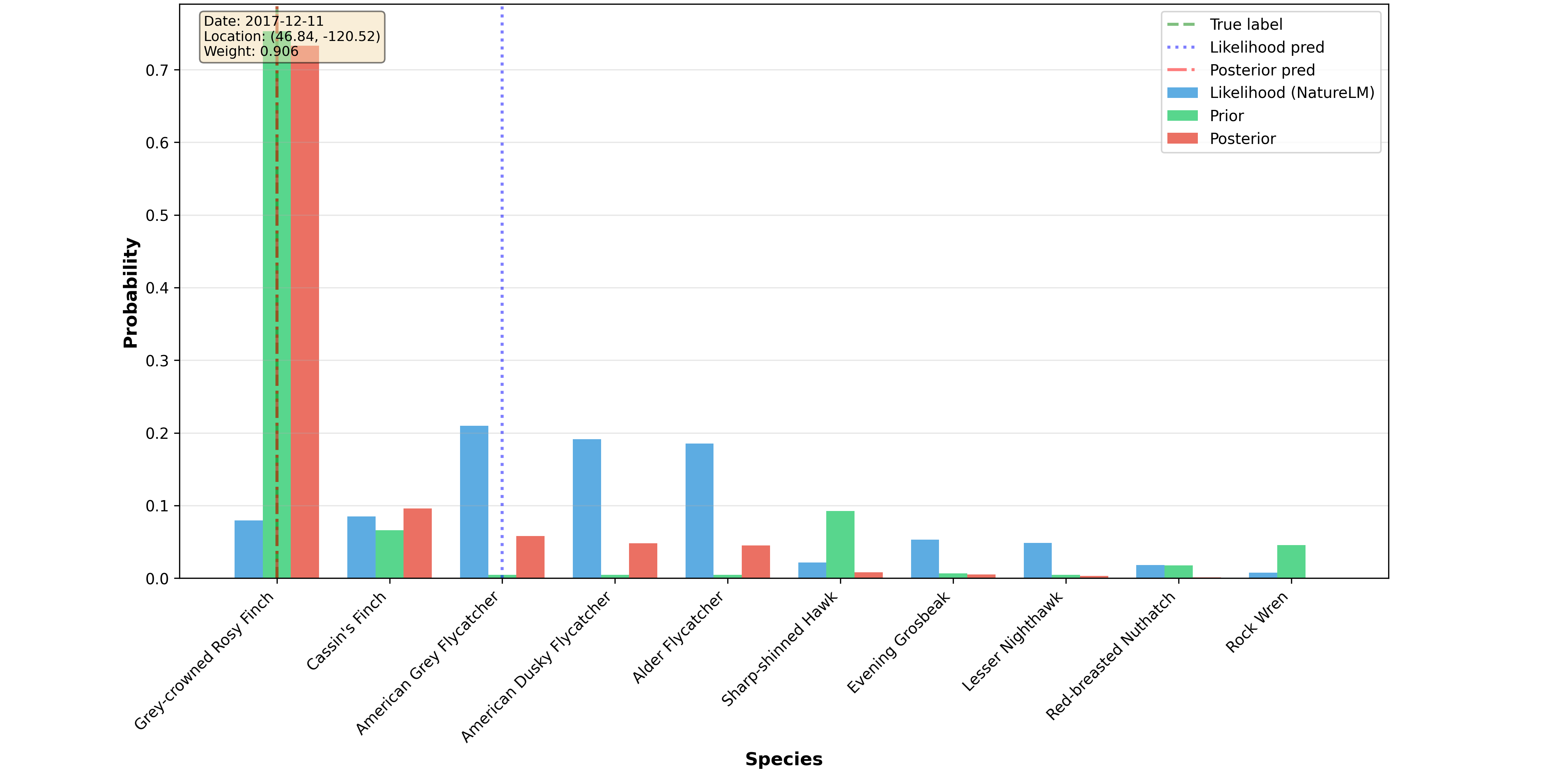}\\[0.5em]
  \includegraphics[width=\linewidth]{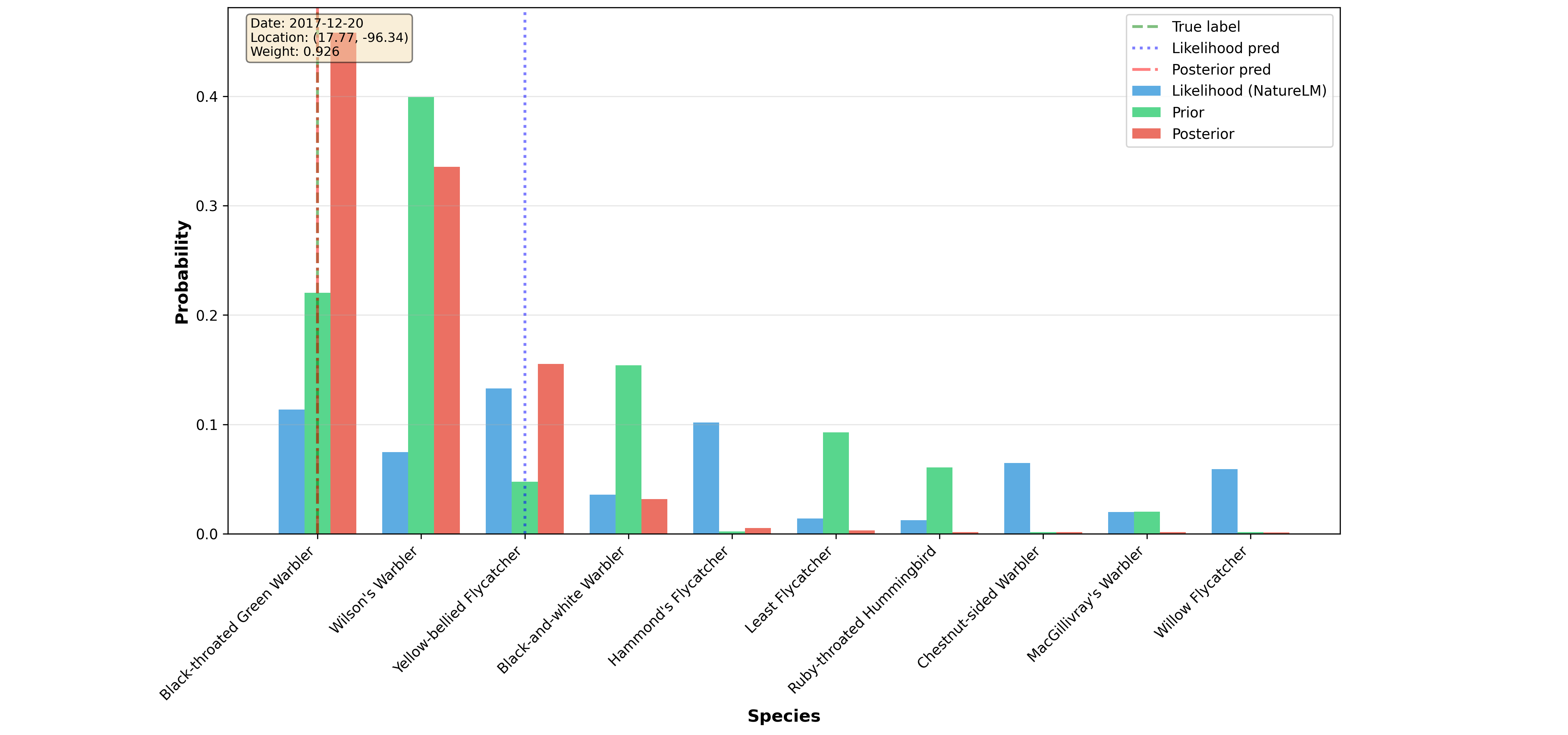}
  \caption{
  \textbf{Motivating examples of adaptive fusion at test time.}
  \textbf{Top:} The acoustic model assigns low probability to the true class (Grey-crowned Rosy Finch), ranking several acoustically similar species higher. The spatiotemporal model is highly confident and geographically specific, and the fused posterior correctly recovers the prior’s top prediction with a high gating weight.
  \textbf{Bottom:} Neither the acoustic model (top prediction: Yellow-bellied Flycatcher) nor the spatiotemporal prior (diffuse over multiple warblers) is correct in isolation. However, adaptive fusion suppresses the acoustically dominant but ecologically implausible class, yielding the correct posterior prediction (Black-throated Green Warbler).
  In both cases, the learned gating weight differs substantially across inputs, illustrating why fixed-weight fusion would fail to recover these outcomes.
  }
  \label{fig:qualitative_examples}
\end{figure*}

To complement the aggregate results, we present two representative test-time examples in Fig.~\ref{fig:qualitative_examples} that illustrate how FINCH adaptively balances acoustic and spatiotemporal evidence.
In the first example, the audio-only classifier assigns low probability to the true species, ranking several acoustically similar alternatives higher, while the spatiotemporal model is highly confident and geographically specific. FINCH assigns a large gating weight and recovers the correct class by amplifying reliable contextual evidence.
In the second example, neither the acoustic model nor the spatiotemporal prior is correct in isolation: the prior is diffuse across multiple species, and the audio model is confidently wrong. Here, adaptive fusion suppresses the acoustically dominant but ecologically implausible class, yielding a correct posterior prediction despite the failure of both marginal predictors.
Importantly, the learned gating weights differ substantially across the two cases, demonstrating that a fixed-weight fusion would be unable to recover both outcomes. These examples highlight the role of adaptive reliability estimation in mediating error trade-offs and align with the conditional-independence analysis in Appendix~A2.

%\input{tables/ablation_1}

%%%%%%%%%%%%%%%%%%%%%%%%%%%%%%%%%%%%%%%%%%%%%%%%%%%%%%%%%%%%%%%%%%%%%%%%%%%%%%%
% 6 DISCUSSION
%%%%%%%%%%%%%%%%%%%%%%%%%%%%%%%%%%%%%%%%%%%%%%%%%%%%%%%%%%%%%%%%%%%%%%%%%%%%%%%

\section{Discussion}
\label{sec:discussion}

This work demonstrates that adaptive log-linear fusion of approximately conditionally independent evidence sources can yield consistent improvements over both standalone predictors and fixed-weight fusion schemes.
Across large-scale bioacoustic benchmarks, FINCH outperforms strong audio-only baselines on evaluated subsets, demonstrating the effectiveness of adaptive evidence fusion under weak and heterogeneous contextual signals.

These gains arise from selectively integrating contextual evidence on a per-sample basis rather than enforcing a global calibration between modalities.

A key property of the proposed framework is its decision-theoretic safety.
Under this formulation, the audio-only classifier serves as an explicit fallback within the fusion family.
When contextual evidence is uninformative or misleading, the learned gating function is encouraged to suppress its influence, reducing the risk of performance degradation relative to the likelihood model.
This behavior is reflected empirically in both aggregate metrics and qualitative examples, where adaptive fusion corrects confident audio errors when context is reliable and avoids pathological domination when it is not.

The design of the gating mechanism involves several modeling choices that we treat pragmatically rather than exhaustively.
In this work, the gating network operates on a compact set of uncertainty- and confidence-based summary statistics derived from the predictive distributions.
Alternative feature constructions or more expressive gating architectures could be explored, but our results indicate that even a lightweight model suffices to capture substantial heterogeneity in contextual reliability.

We also note that the spatiotemporal prior used in this study is relatively weak in isolation.
With $264$ classes, random guessing yields an accuracy of approximately $0.38\%$, whereas the prior achieves roughly $3\%$ accuracy.
This gap confirms that the observed performance improvements stem from selective integration rather than reliance on context alone.
Stronger ecological models could further amplify the benefits of adaptive fusion, but the current results suggest that even coarse contextual signals can meaningfully improve predictions when integrated safely.

Both the CBI dataset and the spatiotemporal prior are derived from the broader eBird ecosystem, which raises natural questions about potential data entanglement.
Importantly, the AdaSTEM priors are trained on independent citizen-science checklists aggregated over large spatial and temporal windows and do not incorporate the audio recordings or competition splits used in CBI.
Moreover, the prior achieves very low standalone accuracy, suggesting that it functions as a weak ecological feasibility signal rather than a proxy for species labels.

An important limitation concerns false positives.
While contextual information can suppress ecologically implausible predictions, the logistic regression--based prior employed here interpolates across space and time, assigning nonzero probability to regions with sparse or no observations.
As a result, false positives are reduced but not eliminated, and stricter feasibility constraints may trade off robustness under dataset shift.

Finally, the conditional independence assumption underlying log-linear fusion is inherently strong and does not hold exactly in real-world data.
We emphasize that conditional independence is used here as a modeling motivation rather than a strict assumption.
Our empirical analysis indicates that residual dependence between audio features and spatiotemporal context exists but is weak and heterogeneous across classes.
This regime aligns with the intended operating conditions of FINCH: independence holds sufficiently well to motivate multiplicative fusion, while adaptive gating mitigates deviations without requiring explicit modeling of audio--context interactions.
When conditional dependence is strong, fully joint models may be preferable, but they incur substantially higher training and data requirements.

More broadly, FINCH is best viewed as a general fusion framework rather than a bioacoustic-specific solution.
The formulation applies to any setting in which multiple discriminative predictors provide complementary, approximately independent evidence about a shared target, and extending this approach to other domains remains a promising direction for future work.
\FloatBarrier
% TODO: interpret results, link back to theory, what the gating learns,
% which contexts are reliable, where priors are brittle.

%%%%%%%%%%%%%%%%%%%%%%%%%%%%%%%%%%%%%%%%%%%%%%%%%%%%%%%%%%%%%%%%%%%%%%%%%%%%%%%
% 7 LIMITATIONS
%%%%%%%%%%%%%%%%%%%%%%%%%%%%%%%%%%%%%%%%%%%%%%%%%%%%%%%%%%%%%%%%%%%%%%%%%%%%%%%

% TODO: dependence on prior quality, dataset shift, calibration issues,
% geographic bias, eBird/observation bias, etc.

%%%%%%%%%%%%%%%%%%%%%%%%%%%%%%%%%%%%%%%%%%%%%%%%%%%%%%%%%%%%%%%%%%%%%%%%%%%%%%%
% REQUIRED UNNUMBERED SECTIONS (ICML template expectations)
%%%%%%%%%%%%%%%%%%%%%%%%%%%%%%%%%%%%%%%%%%%%%%%%%%%%%%%%%%%%%%%%%%%%%%%%%%%%%%%
\section*{Acknowledgments}
% NOTE: Do NOT include in anonymous submission.
% (Leave empty or with a placeholder comment.)
% TODO (camera-ready): funding, collaborators, etc.

\section*{Impact Statement}

This work contributes a general framework for adaptively fusing multiple sources of probabilistic evidence under approximate conditional independence. While our experiments focus on bioacoustic species classification, the proposed method is applicable to a wide range of domains in which heterogeneous predictors provide complementary information.

In ecological monitoring and biodiversity assessment, improved integration of acoustic and contextual information may support more accurate large-scale analysis of species distributions and temporal trends. At the same time, predictions produced by automated systems should not be treated as definitive evidence of species presence without appropriate human validation, particularly in conservation or policy-relevant settings.

More broadly, the proposed fusion mechanism is designed to be robust to unreliable or misleading contextual signals by bounding their influence and preserving a safe fallback to primary evidence. This design choice helps mitigate risks associated with overconfident or biased auxiliary models when deploying decision-support systems in real-world environments.

% REQUIRED by ICML (even in submission).
% Provide a concise statement aligned with your work's ethical and societal impacts.
% TODO: 1--2 paragraphs on ecological monitoring, biases, misuse considerations, mitigation.

%%%%%%%%%%%%%%%%%%%%%%%%%%%%%%%%%%%%%%%%%%%%%%%%%%%%%%%%%%%%%%%%%%%%%%%%%%%%%%%
% REFERENCES
%%%%%%%%%%%%%%%%%%%%%%%%%%%%%%%%%%%%%%%%%%%%%%%%%%%%%%%%%%%%%%%%%%%%%%%%%%%%%%%
\bibliography{bibliography}
\bibliographystyle{icml2026}

%%%%%%%%%%%%%%%%%%%%%%%%%%%%%%%%%%%%%%%%%%%%%%%%%%%%%%%%%%%%%%%%%%%%%%%%%%%%%%%
% APPENDIX (optional; unlimited pages; keep if you anticipate extra tables/derivations)
%%%%%%%%%%%%%%%%%%%%%%%%%%%%%%%%%%%%%%%%%%%%%%%%%%%%%%%%%%%%%%%%%%%%%%%%%%%%%%%
\newpage
\appendix
\onecolumn

\paragraph{Appendix A1: Additional Results}

We report additional experiments that complement the main results and further characterize the behavior of FINCH under alternative contextual models and fusion strategies.

\noindent\textbf{BirdSet learned contextual prior.}
BirdSet does not provide access to an external ecological prior such as eBird Status \& Trends.
For BirdSet experiments, we therefore learn a contextual predictor $p_\psi(y\mid s)$ directly from spatiotemporal metadata.
Specifically, we encode metadata using the same feature representation as in Section~3.3,
\[
f_{\text{meta}}(s)=\Big[\sin(2\pi d/365),\cos(2\pi d/365),\sin(2\pi h/24),\cos(2\pi h/24),\tfrac{\text{lat}}{90},\tfrac{\text{lon}}{180}\Big],
\]
and train a small multilayer perceptron (MLP) on the BirdSet training split to predict species labels from $f_{\text{meta}}(s)$ alone.
After training, this model is treated as the contextual predictor $p_\psi(y\mid s)$ and frozen during FINCH fusion training, matching the setting of independently-trained experts.

Table~\ref{tab:benchmark} reports results on BirdSet subsets across multiple evaluation metrics.
Although the metadata-only prior is weak in isolation, adaptive fusion enables FINCH to match or exceed strong audio-only baselines on several subsets, particularly in Top-1 accuracy.
These results indicate that FINCH remains effective even when contextual information is learned from limited supervision and exhibits substantial heterogeneity across datasets.

\begin{table*}[t]
\centering
\caption{Model results on large-scale bioacoustic classification datasets: BirdSet \cite{rauch2025birdsetlargescaledatasetaudio}, CBI \cite{cbi_data}. (Method) ``Pre": pre-trained classification head, ``FT": full fine-tuning, ``LP": linear probe, ``PP": prototypical probe, ``0": zero-shot. (BirdSet) Following along with Perch2.0 \cite{perchiipointzero}, we report area under ROC (AU-ROC) and class-mean average precision (cmAP). (BEANS) Similarly, here we record model accuracy (Acc) and mean AP (mAP). For CBI, FINCH uses the external AdaSTEM prior; for BirdSet, FINCH uses a learned metadata-only MLP prior trained on the BirdSet training split. Best results highlighted in \textbf{bold}. Second best results \underline{underlined}.}
\label{tab:benchmark}
\begin{tabular}{lcccc cc}
\toprule
 & \multicolumn{4}{c}{\textbf{BirdSet}} & \multicolumn{2}{c}{\textbf{CBI}} \\
 & Method & AUROC & cmAP & Acc & Method & Acc \\
\midrule
Audio ProtoPNet-5 & Pre & 0.896 & 0.423 & 0.623 & --  & -- \\
BirdMAE-L & FT & 0.886 & \textbf{0.440} & 0.601 & -- & -- \\
BirdMAE-L & PP & 0.886 & 0.409 & 0.521 & -- & -- \\
AVES-Bio & -- & -- & -- & -- & FT & 0.598 \\
BirdNet & -- & -- & -- & -- & FT & 0.702 \\
BioLingual & -- & -- & -- & -- & FT & 0.744 \\
NatureLM-Audio & -- & -- & -- & -- & 0 & 0.778 \\
\midrule
Perch 2.0 - Peak-select & Pre & 0.907 & 0.430 & 0.619 & LP & 0.785 \\
 & & & & & PP & 0.789 \\
Perch 2.0 - Random & Pre & \textbf{0.908} & 0.431 & \textbf{0.665} & LP & 0.792 \\
\midrule
Audio-only ($p_\theta(y\mid x)$) & FT & 0.833 & 0.261 & 0.636 & LP & \underline{0.806} \\
Prior-only ($p_\psi(y\mid s)$) & LP & 0.466 & 0.027 & 0.461 & LP & 0.030 \\
FINCH (ours) & FT & 0.832 & 0.260 & \underline{0.636} & LP & \textbf{0.826} \\
\bottomrule
\end{tabular}
\end{table*}
\medskip
\noindent\textbf{Fixed-weight versus adaptive fusion on CBI.}
We further analyze the effect of fixed fusion weights when combining audio predictions with the eBird spatiotemporal prior on a subset of the CBI dataset (Table~\ref{tab:ablation}).
As the fusion weight $\omega$ increases, performance degrades monotonically, reflecting the low informativeness and high entropy of the prior when applied indiscriminately.
In contrast, adaptive fusion substantially outperforms all fixed-weight settings by selectively incorporating contextual evidence only when it is informative.
This experiment highlights a key failure mode of global fusion weights and motivates the use of input-dependent reliability estimation.

\begin{table}[hbpt]
  \centering
  \begin{tabular}{lcc}
    \toprule
    \(\omega\) & Acc $\uparrow$ & mAP $\uparrow$ \\
    \midrule
    0.0 & 64.28 & 0.333 \\
    0.2 & 63.91 & 0.328 \\
    0.4 & 64.12 & 0.332 \\
    0.8 & 60.22 & 0.325 \\
    1.6 & 59.76 & 0.316 \\
    2.0 & 57.72 & 0.311 \\
    Adaptive & \textbf{66.10} & \textbf{0.352} \\
    \bottomrule
  \end{tabular}
  \vspace{10pt}
  \caption{Ablation study for different values of \(\omega\). We report accuracy and mAP for models trained on 1000 samples from CBI. Fixed \(\omega\) replaces the gating network in Eq.~\ref{eq:fusion} with a scalar. (Adaptive) learns to dynamically weight feature evidence.}
  \label{tab:ablation}
\end{table}
\FloatBarrier

\paragraph{Appendix A2: Decision-Theoretical Safety of Gated Fusion}

We show that the proposed gated fusion mechanism is decision-theoretically safe in the sense of \emph{risk containment}: it admits the audio-only classifier as a recoverable special case and therefore cannot be forced to rely on misleading contextual information.

Let $\ell(y,\hat{p}) = -\log \hat{p}(y)$ denote the log-loss. Consider the unnormalized fused score
\[
\log \tilde{p}_\omega(y\mid x,s)
=
\log p_\theta(y\mid x) + \omega(x,s)\,\log p_\psi(y\mid s),
\qquad \omega(x,s)\ge 0,
\]
and define the normalized predictor $p_\omega(y\mid x,s)$ by applying a softmax over $y$.

The key structural property of this formulation is \emph{recoverability}: for any input $(x,s)$,
\[
\omega(x,s)=0
\quad\Longrightarrow\quad
p_\omega(y\mid x,s)=p_\theta(y\mid x).
\]
Thus, the audio-only classifier is always contained within the hypothesis class induced by gated fusion.

As a consequence, for any data-generating distribution and any loss function evaluated pointwise in $(x,s)$, the optimal choice of $\omega(x,s)$ satisfies
\[
\inf_{\omega(x,s)\ge 0}
\;\mathbb{E}\!\left[\ell\big(y,p_\omega(\cdot\mid x,s)\big)\right]
\;\le\;
\mathbb{E}\!\left[\ell\big(y,p_\theta(\cdot\mid x)\big)\right].
\]
In particular, whenever the spatiotemporal predictor $p_\psi(y\mid s)$ is uninformative or misleading for a given input, the choice $\omega(x,s)=0$ achieves the same expected risk as the audio-only model.

Conversely, if contextual information provides non-negative information gain for a subset of inputs, then there exist inputs $(x,s)$ for which $\omega(x,s)>0$ strictly improves expected log-loss relative to the audio-only baseline. The adaptive gating network learns a data-driven approximation to this selective incorporation rule.

Thus, adaptive gated fusion strictly generalizes both audio-only and fixed-weight fusion while preserving a safe fallback.

\paragraph{Appendix A3: When Log-Linear Fusion is Exact (Conditional Independence Case)}

We justify the log-linear fusion rule and clarify its relationship to joint audio--context models. This analysis is intended to clarify the conditions under which log-linear fusion is theoretically exact; it does not assume that these conditions hold universally in practice.

Assume that audio features $x$ and spatiotemporal context $s$ are conditionally independent given the class label $y$, i.e.,
\[
p(x,s\mid y) = p(x\mid y)\,p(s\mid y).
\]
By Bayes' rule, the posterior satisfies
\[
p(y\mid x,s)
=
\frac{p(x,s\mid y)\,p(y)}{p(x,s)}
\propto
p(x\mid y)\,p(s\mid y)\,p(y).
\]
Rewriting this expression in terms of the available predictors,
\[
p(y\mid x) = \frac{p(x\mid y)p(y)}{p(x)}, 
\qquad
p(y\mid s) = \frac{p(s\mid y)p(y)}{p(s)},
\]
we obtain
\[
p(y\mid x,s)
\propto
\frac{p(y\mid x)\,p(y\mid s)}{p(y)}.
\]
Equivalently, up to an additive constant independent of $y$,
\[
\log p(y\mid x,s)
=
\log p(y\mid x)
+
\log p(y\mid s)
-
\log p(y)
+
\text{const}.
\]

\medskip
\noindent\textbf{Effect of Conditional Dependence.}
If conditional independence is violated, the joint distribution can be written as
\[
p(x,s\mid y)
=
p(x\mid y)\,p(s\mid y)\,\kappa_y(x,s),
\qquad
\kappa_y(x,s)
:=
\frac{p(x,s\mid y)}{p(x\mid y)p(s\mid y)}.
\]
The posterior becomes
\[
p(y\mid x,s)
\propto
\frac{p(y\mid x)\,p(y\mid s)}{p(y)}
\;\kappa_y(x,s),
\]
or, in log form (up to a class-independent constant),
\[
\log p(y\mid x,s)
=
\log p(y\mid x)
+
\log p(y\mid s)
-
\log p(y)
+
\log \kappa_y(x,s)
+
\text{const}.
\]
Crucially, $\log \kappa_y(x,s)$ depends explicitly on $y$ and therefore cannot, in general, be represented by a class-agnostic scalar reweighting of $\log p(y\mid s)$.

\medskip
\noindent\textbf{Implications for Model Design.}
When conditional dependence is strong, accurately capturing $\kappa_y(x,s)$ requires a joint predictor that ingests $(x,s)$ together. Such joint models are strictly more expressive but require paired supervision and explicit learning of audio--context interactions.

In contrast, when conditional independence approximately holds, log-linear fusion of independently trained predictors is sufficient and (up to normalization) optimal. FINCH is motivated by this regime: it preserves the posterior form implied by independence while introducing adaptive, reliability-aware weighting to account for model mismatch and heterogeneous uncertainty.

\paragraph{Appendix A4: Empirical Test of Conditional Dependence}

The theoretical justification above relies on a conditional independence assumption. We now empirically assess the extent to which this assumption holds in practice.

Fixing a class label $y$, we test whether audio embeddings $e(x)$ contain information about spatiotemporal context statistics $s$ by training a linear predictor
\[
\hat{s} = f(e(x))
\]
to predict spatiotemporal priors from audio embeddings alone. As a baseline, we compare against a constant predictor that always outputs the mean prior.

Across $43$ species with sufficient samples, the linear predictor does not consistently outperform the baseline on held-out data. While a permutation test indicates that the observed differences are unlikely to arise purely by chance ($p<10^{-3}$), effect sizes are small (Cohen's $d=-0.261$), and explained variance is negligible or negative (mean $R^2=-0.34$). Only $16.3\%$ of classes exhibit positive improvement over the baseline, and fewer than $12\%$ achieve $R^2>0.05$.

Taken together, these results suggest that audio and spatiotemporal information are not strictly conditionally independent, but their dependence is weak and heterogeneous across classes. These findings support the use of adaptive gating: while conditional independence is not exact, deviations are sufficiently weak and variable that a bounded, sample-dependent fusion mechanism is preferable to fixed or fully joint models.

\paragraph{Appendix A5: Spatiotemporal Prior Details}

The spatiotemporal prior is based on weekly relative abundance estimates produced by the eBird Status and Trends project. These estimates are generated using Adaptive Spatio-Temporal Exploratory Models (AdaSTEM), an ensemble of localized regression models designed to account for heterogeneous sampling effort and detection bias in large-scale citizen-science data \cite{fink2010spatiotemporal,Fink2014,Jonhston2015}. The abundance maps are provided on a global $27\,\mathrm{km} \times 27\,\mathrm{km}$ grid at weekly resolution. Each abundance value represents the expected number of individuals detected by an expert observer at a specific location and time. Raw values are stored as unsigned integers in the range $[0,255]$ and linearly rescaled to $[0,1]$.

To avoid introducing a training-time bottleneck, we precompute a spatiotemporal prior lookup table for all training samples. For each audio clip with metadata $(\text{lat}_i, \text{lon}_i, \text{date}_i)$, we query the corresponding weekly abundance prior $p(y_j \mid \text{lat}_i, \text{lon}_i, \text{date}_i)$ for all species $y_j$. The resulting matrix of shape $(N_{\text{samples}}, N_{\text{species}})$ is stored in an HDF5 file and indexed by sample ID. During training, batches retrieve prior values via indexed lookup, yielding an average access time of $0.22\,\mu$s, compared to approximately $0.01$s for on-the-fly querying. At inference time, priors are queried directly from the abundance maps, as latency is not performance-critical.

\paragraph{Appendix A6: Training Configuration and Optimization Details}

All training stages share the same optimization and training configuration unless otherwise noted. Optimization is performed using the AdamW optimizer with learning rate $10^{-3}$ and weight decay $10^{-2}$. Models are trained for 30 epochs with a batch size of 96.

We construct a stratified validation split comprising 10\% of the training data, stratified by species label. Model selection is performed based on validation accuracy.

A cosine learning rate schedule with linear warmup over the first 10\% of training steps is used. Training is performed on GPU using mixed-precision arithmetic (bfloat16).

Checkpoints are saved after every epoch. For each training stage, the checkpoint achieving the highest validation accuracy is retained for evaluation.

Stage~2 is initialized from the best-performing Stage~1 checkpoint, and Stage~3 is initialized from the corresponding Stage~2 checkpoint. The adaptive gating network is initialized such that its output matches the fixed scalar fusion weight learned in Stage~2. This staged initialization stabilizes training.

Throughout all stages, the audio encoder remains frozen. In Stages~2 and~3, the audio classification head is also frozen, and only fusion parameters are updated.

All experiments are implemented in PyTorch. Random seeds are fixed across runs to ensure reproducibility. Unless explicitly stated, no additional data augmentation is applied beyond that inherent to the underlying audio encoder.

\begin{table}[h]
\centering
\caption{Summary of fusion model architectures and training configuration. Parameter counts assume $C=264$ species.}
\label{tab:training_setup}
% \resizebox{\columnwidth}{!}{
\begin{tabular}{lcc}
    \toprule
    \textbf{Component} & \textbf{Stage A} & \textbf{Stage B} \\
    \midrule
    Audio Encoder & Frozen & Frozen \\
    Encoder Architecture & \multicolumn{2}{c}{BEATs (NatureLM-Audio)} \\
    Encoder Output Dim. & \multicolumn{2}{c}{4096} \\
    \midrule
    Audio Classifier & $4096 \times C + C$ & $4096 \times C + C$ \\
    Temperature Parameter & 1 & 1 \\
    Epsilon Parameter & 1 & 1 \\
    Scalar Weight & 1 & -- \\
    Gating Network & -- & 2,945 \\
    \midrule
    Total Trainable Params & $\approx 1.08$M & $\approx 1.09$M \\
    \midrule
    Optimizer & \multicolumn{2}{c}{AdamW} \\
    Learning Rate & \multicolumn{2}{c}{$1 \times 10^{-3}$} \\
    Batch Size & \multicolumn{2}{c}{96} \\
    Epochs & \multicolumn{2}{c}{30} \\
    LR Schedule & \multicolumn{2}{c}{Cosine w/ warmup} \\
    Precision & \multicolumn{2}{c}{bfloat16} \\
    \bottomrule
\end{tabular}
% }
\end{table}

\end{document}